\documentclass[journal=jpcbfk,manuscript=article]{achemso}

\usepackage{caption}
\usepackage{sectsty}
\sectionfont{\MakeUppercase}

\usepackage{amsmath,amssymb}

\usepackage{graphicx}
\usepackage{hyperref}

\usepackage{algorithm}
\usepackage[noend]{algpseudocode}
\makeatletter
\def\BState{\State\hskip-\ALG@thistlm}
\makeatother

\usepackage{color}

\title{Quantitative Kinetic Models from Intravital Microcopy: A Case Study Using Hepatic Transport}

\author{Meysam Tavakoli}
\affiliation{Dept. of Physics, Indiana University-Purdue University, Indianapolis, IN, USA}

\author{Konstantinos Tsekouras}
\affiliation{Dept. of Physics, Arizona State University, Tempe, AZ, USA}

\author{Richard Day}
\affiliation{Dept. of Cellular and Integrative Physiology, Indiana University School of Medicine, Indianapolis, IN, USA}

\author{Kenneth W. Dunn}
\affiliation{Dept. of Medicine and Biochemistry, Indiana University School of Medicine, Indianapolis, IN, USA}

\author{Steve Press{\'e}}
\affiliation{Center for Biological Physics, Department of Physics,\\Arizona State University, Tempe, AZ 85287, USA
and\\
School of Molecular Sciences,\\Arizona State University, Tempe, AZ 85287, USA}
\email{spresse@asu.edu}

\date{\today}


\begin{document}

\textbf{To appear in the Journal of Physical Chemistry B}

\maketitle

\begin{abstract}
The liver performs critical physiological functions, including metabolizing and 
removing substances, such as toxins and drugs, from the bloodstream. 
Hepatotoxicity itself is intimately linked to abnormal hepatic transport 
and hepatotoxicity remains the primary reason drugs in development fail 
and approved drugs are withdrawn from the market.
For this reason, we propose to analyze,  across liver compartments, the transport kinetics of fluorescein--a fluorescent marker used as a proxy for drug molecules--using intravital microscopy data.
To resolve the transport kinetics quantitatively from fluorescence data, we account for
the effect that different liver compartments (with different chemical properties) have on 
fluorescein's emission rate. 
To do so, we develop
ordinary differential equation transport models from the data 
where the kinetics are related to the observable fluorescence levels by 
"\textit{measurement parameters}" that vary across different liver compartments. 
On account of the steep non-linearities in the kinetics and stochasticity inherent to the model, 
we infer kinetic and measurement parameters by generalizing the method of 
parameter cascades.
For this application, the method of parameter cascades 
ensures fast and precise parameter estimates from noisy time traces. 
\end{abstract}

\newpage

\section{Introduction}
\label{sec:intro}

\subsection*{Physiological Context of Hepatotoxicity}
Liver transport is a fundamental physiological process whose significance to human health has increased with the proliferation of pharmaceuticals and environmental toxins~\cite{veldhoen2008aryl, brent1993role, osburn2008nrf2}. Since the liver is a primary venue for the clearance of xenobiotics, it is particularly susceptible to drug-induced injury, in a process known as hepatotoxicity~\cite{sturgill1997xenobiotic,  holt2006mechanisms, navarro2006drug}. 
Drug hepatotoxicity is associated with inhibition of hepatic transport ~\cite{russmann2009current, bissell2001drug, stieger2000drug}
through the inhibition of transporters~\cite{dawson2012vitro, morgan2010interference}.  Drug effects on hepatic transporters are also a major cause of drug-drug interactions, compromising drug safety and complicating drug dosing~\cite{giacomini2010membrane}.  Although hepatic side effects are a primary focus of preclinical drug evaluations, drug-induced liver injury affects an estimated million people each year globally, and is the most common cause for withdrawal of drugs from the market~\cite{temple2002safety, lee2003drug}. 

Typically, the effects of a drug on hepatic transport are first evaluated outside animal models such as in 
studies of vesicle preparations or cultured cells~\cite{kostrubsky2003evaluation}. 
While these simplified systems yield accurate kinetic transport parameters, they also have key limitations: 
1) they do not recapitulate the complexity of typical clinical situations, which may include one or more pathological conditions in an individual taking a combination of drugs~\cite{zimmerman1999hepatotoxicity}; and 2) they lack the pharmacokinetic processes that determine drug distributions, confounding prediction of {\it in vivo} drug effects from {\it in vitro} dose-response curves~\cite{bjornsson2003conduct, wu2005predicting}. In other words, they lack the full complexity of {\it in vivo} transport, a non-vectorial process mediated by the simultaneous activity of multiple transporters~\cite{van2010organic, van2012complete}. 

By contrast, laboratory animals, combined with the tools of intravital microscopy (IVM) data~\cite{babbey2012quantitative},
provide the necessary physiological context~\cite{marx2012human}. 
The failure to predict drug transport inside the liver from IVM data, however, highlights fundamental shortcomings in how we exploit the data. 
In principle, the data contains information on
the mechanism of vectorial drug transport involving
different transporters, often with overlapping specificities. Imaging methods also, simultaneously, are poised to provide spatial and temporal resolution 
on how drugs might impact liver transport 
from their point of uptake into hepatocytes, through secretion into the bile, with secretion back into the blood, or flow in the biliary tract~\cite{chan2004abcs, sherlock2008diseases}.

Recently, some studies have used IVM~\cite{babbey2012quantitative, presson2011two, lorenz2012digital} to monitor transport kinetics of sodium fluorescein~\cite{de2011sodium, mor2001identification} and identify the effects of chronic kidney disease on organic anion transport~\cite{ryan2014effects}. 
While rich in structure, the IVM data also presents important challenges toward achieving a complete picture of fluorescein's transport kinetics as it evolves from the liver capillaries (sinusoids) into the cytosol of the hepatocytes (uptake) and then into the bile canaliculi (canalicular secretion), from which they are cleared into the bile.
However, fluorescein's emission is deeply dependent on its local chemical environment. That is, the fluorescence signal from these probes is sensitive to environmental quenching~\cite{blair1986fluorescein, chahal1985metabolism} and fluorescein itself may exist in multiple forms, e.g., glucuronidated form~\cite{blair1986fluorescein}, across liver compartments. 

Thus, in this study, we combine experiments and theory to develop a quantitative method to analyze hepatic transport from fluorescence time measurements using IVM data. 
In particular, we model the kinetics of hepatic transport, in other words the kinetics of transport of fluorescent species, 
using a set of ordinary differential equations (ODEs)~\cite{vanlier2013parameter, kwang2003mathematical, cho2003investigations, swat2004bifurcation, tyson2001network, voit2000computational}.
We treat the units of fluorescing species in a particular compartment and their kinetics between liver compartments 
as a hidden (latent) variables and introduce \textit{measurement parameters} to describe 
the relationship between the absolute concentrations and fluorescence intensity in different observed regions.
We calibrate our ODE model, i.e., infer kinetic and measurement  parameters, from noisy
fluorescence time traces obtained from IVM using the method of parameter cascades~\cite{ramsay2007parameter}.

\subsection{Mathematical Methods of ODE Parameter Estimation} \label{Mathematical methods of ODE parameter estimation}

A number of parameter estimation methods exist~\cite{bard1974nonlinear, biegler1986nonlinear, ramsay2006functional, brunel2008parameter, chen2008efficient, conrad2015probability, rosales2004mcmc, siekmann2011mcmc, girolami2011riemann, dondelinger2013ode, calderhead2009accelerating, gupta2016computational} 
some of which we have recently  reviewed~\cite{tavakoli2016single, lee2017unraveling}. 
Here, we adapt the method of parameter cascades~\cite{ramsay2007parameter} which is computationally efficient,  
maintains good numerical efficiency for estimation of ODE parameters from data ~\cite{cao2008estimating}, works for linear or nonlinear dynamics,
straightforwardly estimates measurement parameters
and takes simultaneous advantage of all points in a time trace to perform parameter inference~\cite{wang2014estimating}.
Using this method, ODE solutions are approximated using spline coefficients. 
These coefficients are estimated with penalized smoothing splines with a roughness penalty term.


\section{MATERIALS AND METHODS}
 \label{Materials and Method}

\subsection{Experimental Methods}

Here we used quantitative IVM data for transport in the liver of rats with 5/6th nephrectomy (5/6N)~\cite{laouari2001two, naud2007effects} where hepatic drug transport is impacted by chronic kidney disease~\cite{naud2007effects, ryan2014effects}. 
This data was previously published and information on 5/6N rat models and IVM data collection is detailed in Refs.~\cite{ryan2014effects} and~\cite{babbey2012quantitative}.

\subsection{ODE Model and Parameter Estimation}

We begin with a set of ODEs describing the evolution in time, $t$, of a species vector, $\textbf{x}$, of length $m$ whose elements are units of fluorescence in a particular compartment

\begin{equation}
\label{basicODE}
\dot{\textbf{x}}(t)={\textbf{f}}(\textbf{x},t|{\boldsymbol\theta}).
\end{equation}

In particular, the species coincide with different chemical forms of fluorescein, i.e., modified by being glucuronidated~\cite{blair1986fluorescein}, and unmodified forms in each compartment. 
The vector ${ \boldsymbol\theta}$ contains parameters (kinetic rates describing transport parameters between liver compartments) 
whose values are {\it a priori} unknown. 
The vector, $\textbf{x}$ itself is not directly observed. Rather, we supplement the dynamical model above with the following measurement model

\begin{equation}
\textbf{y}(t)=\textbf{H}\textbf{x}(t)+{\boldsymbol\epsilon}(t)
\label{measurement}
\end{equation}
where $\textbf{y}(t)$ is a noisy vector of length $r$ describing total units of fluorescence measurements in each  compartment at time $t$ and ${\boldsymbol\epsilon}$ describes the noise associated with the measurement assumed white noise with zero mean and finite variance ($\sigma^2$). $\textbf{H}$ is an $m \times n$ measurement matrix in the measurement equation, Eq.~(\ref{measurement}), which relates the state to the measurement output similar to an equivalent matrix appearing in Kalman filtering~\cite{welch1995introduction}.
Each element of $\textbf{H}$ is called a \textit{measurement parameter}. 
For example, a measurement parameter of $0.15$ indicates that only $15\%$ of the substrate is observed, while the remaining $0.85$ is unobserved. 
It naturally follows that all measurement parameters take values between $[0,1]$. 
We define a vector of  all \textit{measurement} and \textit{kinetic} parameters, called \textit{structural parameters}, 
${\boldsymbol\theta'} = \{\boldsymbol\theta,\textbf{H}\}$ and assume that the variances associated with the noise are, \textit{a priori}, known. 

Next, we use the method of parameter cascades~\cite{ramsay2007parameter} to learn the parameters from the data.  
To do so, we first approximate the solution, $\textbf{x}(t)$, of the ODE, Eq.~(\ref{basicODE}), with a linear combination of K basis functions, ${\Phi}=\{\varphi_{k}\}$, k=1,..., K, 
as follows
\begin{equation} \label{eq:ODE0}
\widehat{x}_{i}=\sum_{k=1}^{K} c_{ik}\varphi_{k}\Rightarrow \widehat{\textbf{x}}=\textbf{c}{\Phi}
\end{equation}
where $\widehat{\textbf{x}}$ is the approximation of the curve $\textbf{x}$ in terms of our linear expansion.
We use $i$ to iterate over the $m$ species in our model and call the expansion coefficients $c_{ik}$ nuisance parameters. 
The basis functions must themselves approximate the ODE solutions. 
We selected B-splines as these basis functions allow us to appropriately control solution smoothness across time as warranted by the data which serves as input~\cite{ramsay2006functional, cao2008estimating}.
The number of basis functions must be large enough to adequately represent $\textbf{x}$~\cite{zhang2015selection} and the function $\widehat{x}_{i}$ must be learned by optimizing a global objective function that, at once, satisfies the ODE and adequately fits the noisy data~\cite{cao2012modeling}.
We then iterate between two optimization routines until a pre-specified criterion for the global optimum is met. 
In the first optimization step, the nuisance parameters, $\textit{\textbf{c}}$, are estimated using a smoothing ODE-penalized criterion, in a process known as inner optimization. 
Within the inner optimization, the structural parameters are kept fixed and the nuisance parameters are fitted to data by minimizing the following penalized sum of squares.

\begin{equation}
J_{in}(\textbf{c}|{\boldsymbol \theta'},{\boldsymbol \lambda},\textbf{y})=\sum_{i=1}^{n}|y_{i}-\widehat{x}_i|^{2} + \boldsymbol{\lambda} PEN(\widehat{\textbf{x}}).
\label{penalSum0}
\end{equation}

In the inner optimization, the regularization parameter, ${\boldsymbol \lambda}$, controls the trade off between fitting the data and fidelity to the ODEs for each $x_i$. Intuitively, for larger noise, as defined by Eq.~(\ref{measurement}), we need larger ${\boldsymbol \lambda}$ as the data themselves become less reliable.

In the outer optimization step, the structural parameters, ${\boldsymbol\theta'}$, are updated by minimizing the following sum of squared errors between the data $y_i$ and our estimates $\widehat{x}_{i}$

\begin{equation}
J_{out}({\boldsymbol \theta'}|{\boldsymbol \lambda},\textbf{y})=\sum_{i=1}^{n}|y_{i}-\widehat{x}_i |^{2}.
\label{optzouter}
\end{equation}

Here $J_{out}({\boldsymbol \theta'}|{\boldsymbol \lambda},\textbf{y})$ is minimized with respect to ${\boldsymbol\theta'}$ by using the Newton-Raphson method.
The following pseudo-code (further detailed in Supplementary Information Appendix A) sketches this procedure.

\begin{algorithm}[h!]
\caption{}\label{euclid}
\begin{algorithmic}
\Procedure{}{}
\BState Set a value for ${\lambda}$ 
\BState Pick initial values for $\widehat{\textbf{c}},{\widehat{\boldsymbol\theta}'}$
\BState Inner optimization loop:
\State Estimate $\widehat{\textbf{c}}$ by minimizing ${J}_{in}(\widehat{\textbf{c}}'|{\boldsymbol\theta'},{\boldsymbol\lambda})$ 
\State given by Eq.~(\ref{penalSum0}) 
\BState Outer optimization loop:
\State  Estimate $\widehat{\boldsymbol\theta}'$ by minimizing  ${J}_{out}({\boldsymbol\theta'}|\textbf{y},{\boldsymbol\lambda})$ 
\State given by Eq.~(\ref{optzouter})  
\If {  $\widehat{\boldsymbol\theta}'$ are changed (to within some preset precision), ${J}_{in}(\widehat{\textbf{c}}'|{\boldsymbol\theta'},{\boldsymbol\lambda})$ is reoptimized with respect to $\widehat{\textbf{c}}$ } \textbf{goto} \emph{Inner optimization}.
\Else {
\emph{$\widehat{\boldsymbol\theta}'$ = ${\boldsymbol\theta}'$ }}.
\EndIf
\State \textbf{close};
\EndProcedure
\end{algorithmic}
\end{algorithm}

To be clear, we explicitly include measurement parameters among the structural parameters. The ability to incorporate measurement parameters  constitutes an important generalization of the method of cascades to deal with noisy data that was previously suggested~\cite{ramsay2007parameter}. 
We highlight here that the method of cascades is an important, fast and general alternative to extended Kalman filters or other Kalman filter variants~\cite{mobed2016state, jazani2019method, jazani2019alternative}. Kalman filters may solve similar problems to that above but may suffer in the case of pronounced non-linearities in the dynamics, i.e., Eq.~(\ref{basicODE}).
This is especially relevant to us here as we would like our method to hold for a broad range of non-linear dynamics~\cite{lillacci2010parameter, lillacci2012distribution}.

In the Supplementary Information Appendix B, we describe in greater detail how confidence intervals of parameters estimates are determined. 
Briefly, here we mention that if the data are poor or data sets are too small for the number of parameters to be estimated, 
the global objective function may be flat around its maximum and unable to sharply discriminate between different parameter values
(a problem known as "weak identifiability"~\cite{raue2013joining, vanlier2012integrated}). By contrast, "structural unidentifiability" arises when model parameters are not independent and different parameter choices result in equally good fits~\cite{chis2011structural}.

In our case by using a method drawn from Ref.~\cite{walter1997identification}, we identified which structural parameter(s) are unidentifiable 
and input their values from other sources before estimating other parameters. 
To do this, we used an approach proposed in Ref.~\cite{bellman1970structural} detailed in the Supplementary Information Appendix B.

\subsection*{Method Validation} \label{MethodValidation}
To test our method, we validate its performance on systems of increasing complexity 
using sets of simulated (i.e., synthetic) data, where the ground truth is known. 

\begin{figure}[h]
\begin{center}
\hspace{-1.5cm}
     \includegraphics[scale=0.7]{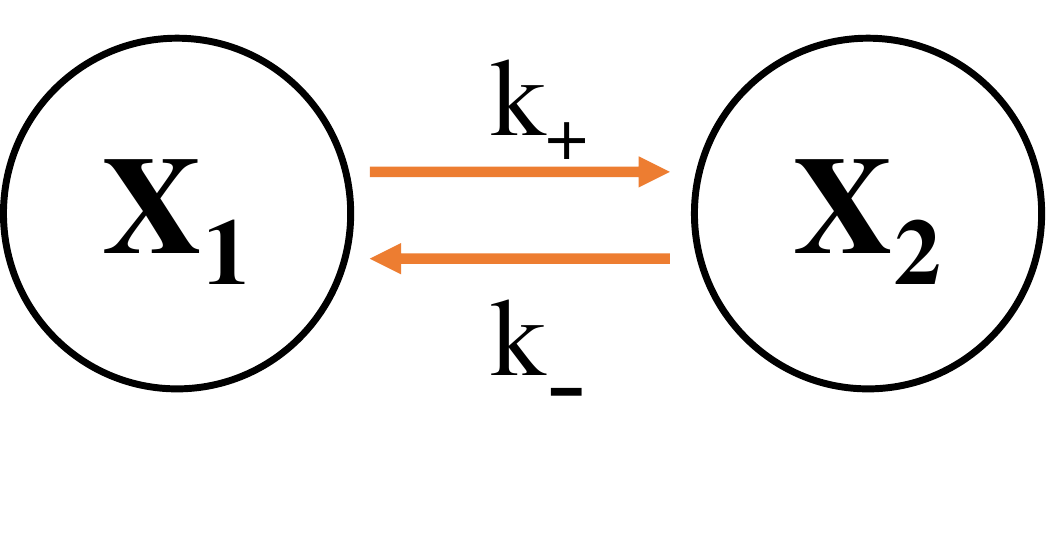}
\caption{{\bf Schematic of Two States Markov Model.} Here, we have two rates, $k_+$ and $k_-$, corresponding to these two states.}
\label{2states}
\end{center}
\vspace{-0.3in}
\end{figure}

{1-	\textit{Two States System}:} 
In the simplest example, we have started with a Two States (compartments or pixels) model whose (Markov) kinetics are determined by two transition rates. Fig.~(\ref{2states}) illustrates this simple two states Markov model. 

The linear ODEs describing this system are

\begin{equation}
\left\{
 \begin{array}{ll}
\dfrac{dX_{1}}{dt}=-k_{+}X_{1}+k_{-}X_{2} \\
\\
\dfrac{dX_{2}}{dt}=k_{+}X_{1}-k_{-}X_{2}
 \end{array}
  \right.
\label{eq2states}
\end{equation}
\\
with measurements

\begin{equation}
\begin{pmatrix} y_{X_{1}} \\ y_{X_{2}} \end{pmatrix} 
= 
\begin{pmatrix} \alpha & 0 \\ 0 & \beta \end{pmatrix}
\begin{pmatrix} X_{1} \\ X_{2} \end{pmatrix} + \begin{pmatrix} \epsilon_1\\ \epsilon_2 \end{pmatrix} 
\label{eqObs2states}
\end{equation}
where ${\boldsymbol\theta'} = {[k_{+},k_{-},\alpha,\beta]}$ being the unknown structural parameter vector.
The mean of both $ \epsilon_1$ and $ \epsilon_2$ is zero and the variance of both is assumed known, i.e., the measurement noise assumed Gaussian is fixed in a pre-calibration step. 
To resolve structural unidentifiability (see the Supplementary Information Appendix B), we must specify either $\alpha$ or $\beta$. For concreteness, we 
presume that from other experiments, it is known that $\alpha=0.5$ and thus we are left with 3 unknown parameters. 

\begin{figure}[h!]
\begin{center}
\hspace{-1.5cm}
     \includegraphics[scale=0.3]{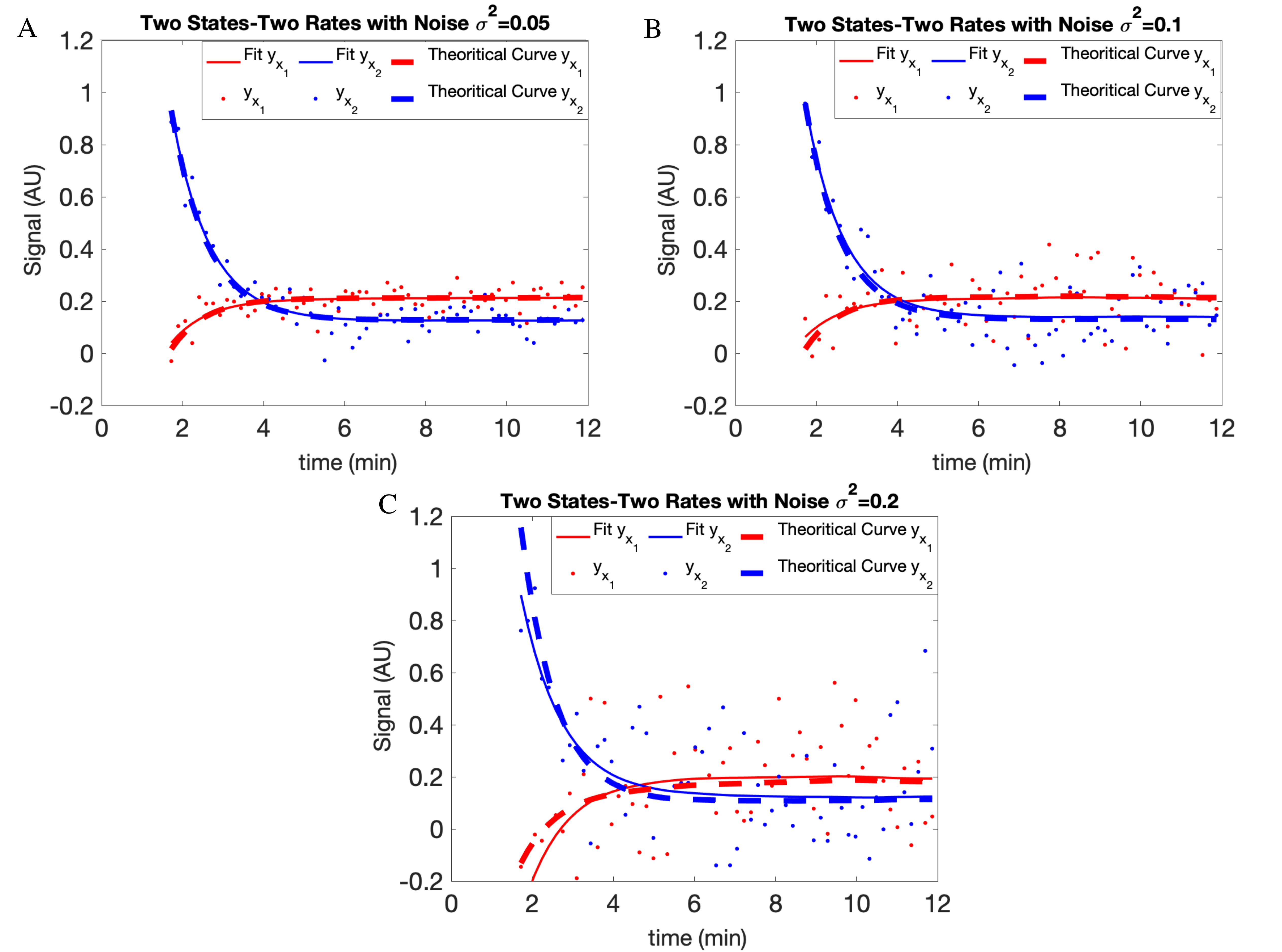}
\caption{{\bf Fits for the Synthetic Model Shown in Fig.~(\ref{2states}).} For each compartment the results  are shown after adding white noise with variance $0.05$ (A), $0.1$ (B) and $0.2$ (C). The dots are generated data points, the dash lines are theoretical curves, and the solid lines are the fits on our data. AU here stands for arbitrary units.}
\label{2statesFit}
\end{center}
\vspace{-0.3in}
\end{figure}

The solutions to Eqs.~(\ref{eq2states}) and (\ref{eqObs2states}) are plotted in Fig.~(\ref{2statesFit}) for parameter values $[0.5, 2, 0.5, 0.3]$ 
and known initial conditions $[y_{X_{1}}(0), y_{X_{2}}(0)]$ = [0, 1].

We also tested the accuracy of our approach by adding white noise with different variances, namely ${0.05, 0.1}$ and ${0.2}$, to our simulated data. 
In Fig.~(\ref{2statesFit}) and Table (\ref{2statesResult}), we show the results of our fitting and parameter estimation.  
\begin{table}[h]
	\centering
	\begin{tabular}{c c c c c}
		\hline\hline
		$k_{+}$ & $k_{-}$ & $\beta$ \\
		\hline\hline
		& & {\bf Without Noise} & &  \\ \hline
		0.5 & 2.0 & 0.3 & True values \\
		0.5000 & 2.0000 & 0.3000 & Estimated values \\
		0.0001 & 0.0002 & 0.0001 & Standard deviation \\
		 \hline
		& & {\bf $\sigma^2 = 0.05$ for the Noise} & &  \\ \hline
		0.5 & 2.0 & 0.3 & True values \\
		0.4555 & 1.9854 & 0.3120 & Estimated values \\
		0.0466 & 0.2809 & 0.0206 & Standard deviation \\ 
		 \hline
		& & {\bf $\sigma^2 = 0.1$ for the Noise} &  &  \\ \hline
		0.5 & 2.0 & 0.3 & True values \\
		0.5202 & 2.2808 & 0.3185 & Estimated values \\
		0.0180 & 0.1434 & 0.0650 & Standard deviation \\
		 \hline
		& & {\bf $\sigma^2 = 0.2$ for the Noise} &  &  \\ \hline
		0.5 & 2.0 & 0.3 & True values \\
		0.5582 & 1.7767 & 0.3381 & Estimated values \\
		0.1134 & 0.3351 & 0.1266 & Standard deviation \\
	\end{tabular}
	\caption{{\bf Two State Results.} The results of parameter estimation for the two state system captured in Fig.~(\ref{2states}) with and without considering noise (the rates are in units of $min^{-1}$). 
The standard deviations reported for the structural parameters are taken from one representative run after determining these parameters over multiple runs of a given synthetic data set.}
	\label{2statesResult}
\end{table}\\

{2-	\textit{Generalized Two States System}:} We continue testing our approach by generalizing the previous example. That is, we have two de-coupled sets of  ODEs for the dynamics whose outputs are coupled by measurement. That is, we have
\begin{equation}
\left\{
 \begin{array}{ll}
\dfrac{dX_{1}}{dt}=-k_{+}X_{1} + k_{-}X_{2} \\ 
\\
\dfrac{dX_{2}}{dt}=k_{+}X_{1}-k_{-}X_{2} \\ 
 \end{array}
  \right.
\label{ODE3states2rates}
\end{equation}

 \begin{equation}
 \left\{
 \begin{array}{ll}
\dfrac{dX'_{1}}{dt}=-k'_{+}X'_{1} + k'_{-}X'_{2} \\ 
\\
\dfrac{dX'_{2}}{dt}=k'_{+}X'_{1}-k'_{-}X'_{2} \\ 
 \end{array}
  \right.
\label{ODE3states2rates1}
 \end{equation}

and 

\begin{equation}
\begin{pmatrix} y_{X_{1}} \\ y_{X_{2}} \end{pmatrix} 
= 
\begin{pmatrix} \alpha & 0 & \alpha' & 0 \\ 0 & \beta & 0 & \beta'\\  \end{pmatrix}
\begin{pmatrix} X_{1} \\ X_{2} \\ X'_{1} \\ X'_{2} \end{pmatrix} + \begin{pmatrix} \epsilon_1\\ \epsilon_2 \end{pmatrix}. 
\label{EqObs3states2rates}
\end{equation}

The above reflects, for example, two different fluorescent species (the primed and unprimed) hopping between two compartments (subscripted one and two).
Measurements on both compartments reveal the total amount of fluorescent material in each compartment but does not discriminate between the primed and unprimed.

\begin{figure}[h]
\begin{center}
\hspace{-1.5cm}
     \includegraphics[scale=0.3]{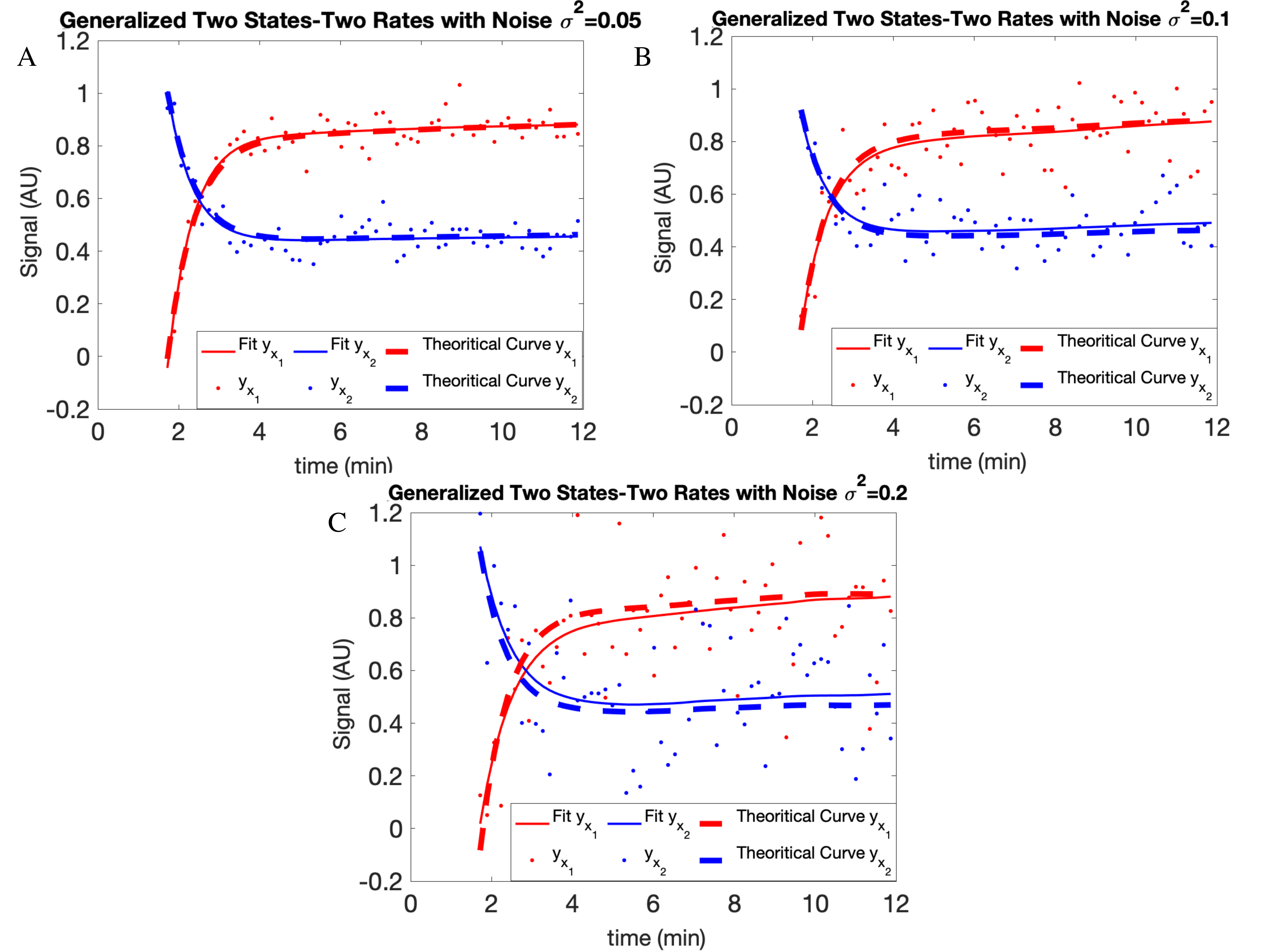}
\caption{{\bf Fits for Generalized Two States Systems.} For each compartment the results of adding white noise with variance $0.05$ (A), $0.1$ (B) and ${0.2}$ (C) have been shown. The dots are generated data points, dash lines are theoretical curves, and the solid lines are the fits on our data.}
\label{3states}
\end{center}
\vspace{-0.3in}
\end{figure}

The structural parameter vector here is ${\boldsymbol \theta'}= [k_+,k_-,k'_+,k'_-,\alpha,\beta,\alpha',\beta']$. To eliminate structural unidentifiability, using the procedure highlighted earlier, we specify 
$\alpha=0.5$ and $\alpha'=0.25$. 
The solutions to Eqs.~(\ref{ODE3states2rates})-(\ref{EqObs3states2rates}) are plotted in Fig.~(\ref{3states}) for parameter values $[0.3, 0.5, 2.0, 4.0, 0.5, 0.5, 0.25, 0.15]$ and initial conditions $[y_{X_1} (0), y_{X_2} (0)]$ = [0, 1]. 
The noise, $\epsilon_1$ and $\epsilon_2$, is treated as we did earlier.

We test the accuracy of our approach by considering white noise with different variances (${0.05, 0.1}$ and ${0.2}$) added to our simulated data. 
In Fig.~(\ref{3states}) and Table~(\ref{GN2states}) we show the results of our fitting and parameter estimation with and without noise.

\begin{table}[h]
	\centering
	\begin{tabular}{c c c c c c c}
		\hline\hline
		$k_{+}$ & $k_{-}$ & $k'_{+}$ & $k'_{-}$ &$\beta$ & $\beta'$ \\
		\hline\hline
		& & {\bf Without Noise} & &  \\ \hline
		0.3 & 0.5 & 2.0 & 4.0 & 0.5 & 0.15 & True values \\
		0.2991 & 0.5000 & 2.0012 & 4.0001 & 0.5000 & 0.1501 & Estimated values \\
		0.0023 & 0.0001 & 0.0016 & 0.0004 & 0.0008 & 0.0005 & Standard deviation \\
		 \hline
		& & {\bf $\sigma^2 = 0.05$ for the Noise} & &  \\ \hline
		0.3 & 0.5 & 2.0 & 4.0 & 0.5 & 0.15 & True values \\
		0.3055 & 0.5101 & 2.0192 & 4.0501 & 0.4938 & 0.1471& Estimated values \\
		0.0234 & 0.0005 & 0.0737 & 0.0403 & 0.0062 & 0.0043 & Standard deviation \\ 
		 \hline
		& & {\bf $\sigma^2 = 0.1$ for the Noise} &  &  \\ \hline
		0.3 & 0.5 & 2.0 & 4.0 & 0.5 & 0.15 & True values \\
		0.2795 & 0.4316 & 1.7738 & 4.2755 & 0.5729 & 0.1421 & Estimated values \\
		0.1064 & 0.1884 & 0.1348 & 0.1229 & 0.1261 & 0.1087 & Standard deviation \\
		 \hline
		& & {\bf $\sigma^2 = 0.2$ for the Noise} &  &  \\ \hline
		0.3 & 0.5 & 2.0 & 4.0 & 0.5 & 0.15 & True values \\
		0.1993 & 0.6909 & 1.7047 & 3.2478 & 0.4247 & 0.1967 & Estimated values \\
		0.1231 & 0.1983 & 0.1801 & 0.1293 & 0.1129 & 0.0915 & Standard deviation \\
	\end{tabular}
	\caption{ {\bf Generalized Two States Results.} The results of parameter estimation for generalized two states with and without considering noise (the rates are in units of $min^{-1}$). 
See Table~(\ref{2statesResult}) for details on the standard deviation.}
	\label{GN2states}
\end{table}

{3- \textit{FitzHugh-Nagumo Model}:} Finally, we tested our method with one of the best known models, developed by FitzHugh~\cite{fitzhugh1961impulses} and Nagumo \textit{et al.}~\cite{nagumo1962active} to examine the behavior of spike potentials in the giant axon of squid neurons. While this model is dissimilar in structure to our  hepatic transport model, the FitzHugh-Nagumo model, shown below, is often used as a benchmark in ODE parameter estimation problems~\cite{ramsay2007parameter, cao2011robust} 
\begin{equation}
\left\{
 \begin{array}{ll}
\dfrac{dV}{dt}=c(V-\dfrac{V^{3}}{3}+R) \\ 
\\
\dfrac{dR}{dt}=\dfrac{1}{c}(V-a+bR). \\ 
 \end{array}
  \right.
\label{FitzHugh-Nagumo}
\end{equation}

This system describes the mutual dependency between voltage across an axon membrane, $V$, and a recovery variable $R$ summarizing outward currents. In this case we setup simulated data for parameter values $[a, b, c]=[0.2, 0.2, 3]$ and initial conditions $[V(0), R(0)]=[-1, 1]$.

In addition to the above, we supplement the dynamical model with a measurement model
\begin{equation}
\begin{pmatrix} y_{V} \\ y_{R} \end{pmatrix} 
= 
\begin{pmatrix} \alpha & 0 \\ 0 & \beta \end{pmatrix}
\begin{pmatrix} V \\ R \end{pmatrix}  + \begin{pmatrix} \epsilon_1\\ \epsilon_2 \end{pmatrix}.
\label{FitzHugh-Nagumo Observation}
\end{equation}

\begin{figure}[h]
\begin{center}
\hspace{-1.5cm}
     \includegraphics[scale=0.3]{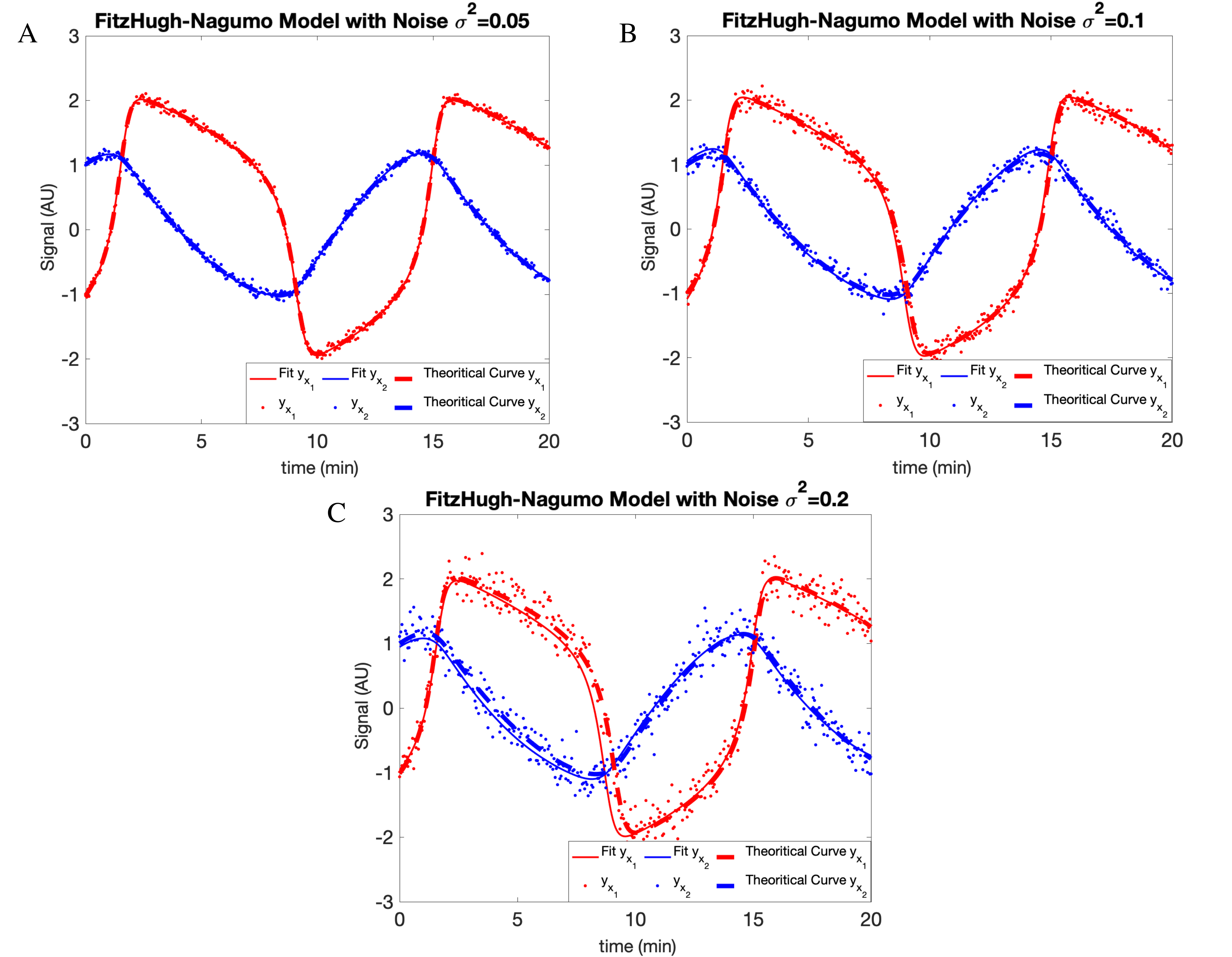}
\caption{{\bf Fit to the FitzHugh-Nagumo Model Supplemented by a Measurement Model.} As before, we consider the following different noise variance levels:  $0.05$ (A), $0.1$ (B) and ${0.2}$ (C). Details in text.}
\label{FitzHugh-NagumoObservResult}
\end{center}
\vspace{-0.3in}
\end{figure}

Thus, the parameters to be determined are now ${\boldsymbol\theta'}= [a,b,c,\alpha,\beta]$. Identifiability demands that we specify either $\alpha$ or $\beta$. 
For this reason, here we set $\alpha$ = 0.5. For synthetic data generated using $[0.2, 0.2, 3.0, 0.5, 0.75]$ and initial conditions $[V(0), R(0)]=[-1, 1]$, the results are shown in Fig.~(\ref{FitzHugh-NagumoObservResult}) and Table~(\ref{FitzHugh-Nagumo Observ Model}).

\begin{table}[h]
	\centering
	\begin{tabular}{c c c c c}
		\hline\hline
		$a$ & $b$ & $c$ & $\beta$\\
		\hline\hline
		& & {\bf Without Noise} & &  \\ \hline
		0.2 & 0.2 & 3 & 0.75 & True values \\
		0.2001 & 0.1999 & 3.0001 & 0.7500 & Estimated values \\
		0.0001 & 0.0013 & 0.0032 & 0.0001 & Standard deviation \\
		 \hline
		& & {\bf $\sigma^2 = 0.05$ for the Noise} & &  \\ \hline
		0.2 & 0.2 & 3 & 0.75 & True values \\
		0.2005 & 0.1995 & 3.0020 & 0.7498 & Estimated values \\
		0.0012 & 0.0016 & 0.0160 & 0.0018 & Standard deviation \\ 
		 \hline
		& & {\bf $\sigma^2 = 0.1$ for the Noise} &  &  \\ \hline
		0.2 & 0.2 & 3 & 0.75 & True values \\
		0.1895 & 0.1841 & 3.046 & 0.7702 & Estimated values \\
		0.0114 & 0.0280 & 0.0243 & 0.0124 & Standard deviation \\
		 \hline
		& & {\bf $\sigma^2 = 0.2$ for the Noise} &  &  \\ \hline
		0.2 & 0.2 & 3 & 0.75 & True values \\
		0.1543 & 0.2317 & 2.9056 & 0.6924 & Estimated values \\
		0.0189 & 0.0372 & 0.0662 & 0.0726 & Standard deviation \\
	\end{tabular}
	\caption{ {\bf FitzHugh-Nagumo Model Results.} The results of parameter estimation for the FitzHugh-Nagumo model obtained 
	by considering increasing noise as before (the rates are in units of $min^{-1}$). See Table~(\ref{2statesResult}) for details on the standard deviation.}
	\label{FitzHugh-Nagumo Observ Model}
\end{table}

As expected across all models, increasing the noise variance level decreases our parameter estimation accuracy and robustness (i.e., error bar). 
The amount by which increased noise reduces the accuracy and robustness of our estimates depends on the model under consideration. So much is clear by comparing, with a variance of $0.2$ for the white noise, the results from Fig.~(\ref{2states}) to those of Fig.~(\ref{FitzHugh-NagumoObservResult}).\\


\section{Results}
\label{sec:res}

\subsection{Full Hepatic Transport Model} \label{Full model of hepatic transport}

We now construct a model of hepatic transport. 
Transport in the liver consists of fluorescein transport between and through sinusoid blood vessels, into the hepatocytes and then into the canaliculi~\cite{babbey2012quantitative}. The data we have collected consists of fluorescence intensity from fluorescein in all three compartments.

To construct our model, we: 1) assume no direct transport between sinusoid and canaliculus; 
and 2) assume only three compartments (sinusoid, hepatocyte and canaliculus). 
In this case, we designate fluorescein species in the sinusoid, hepatocyte and canaliculus as S(t), H(t), and C(t) respectively. In full generality, we also consider back flow from the hepatocyte back into the sinusoid. 

We treat fluorescein in each compartment as a different species with a different measurement parameter since each compartment presents variable quenchers species and concentrations
(e.g., binding proteins reducig fluorescein net emission)~\cite{babbey2012quantitative}.  
What is more, we consider two forms of fluorescein, both unmodified and glucuronidated  
as it is known that the majority of fluorescein is glucuronidated within 30 minutes of intravenous injection~\cite{blair1986fluorescein}. 

\begin{figure}[h]
\begin{center}
\hspace{-1.5cm}
     \includegraphics[scale=0.7]{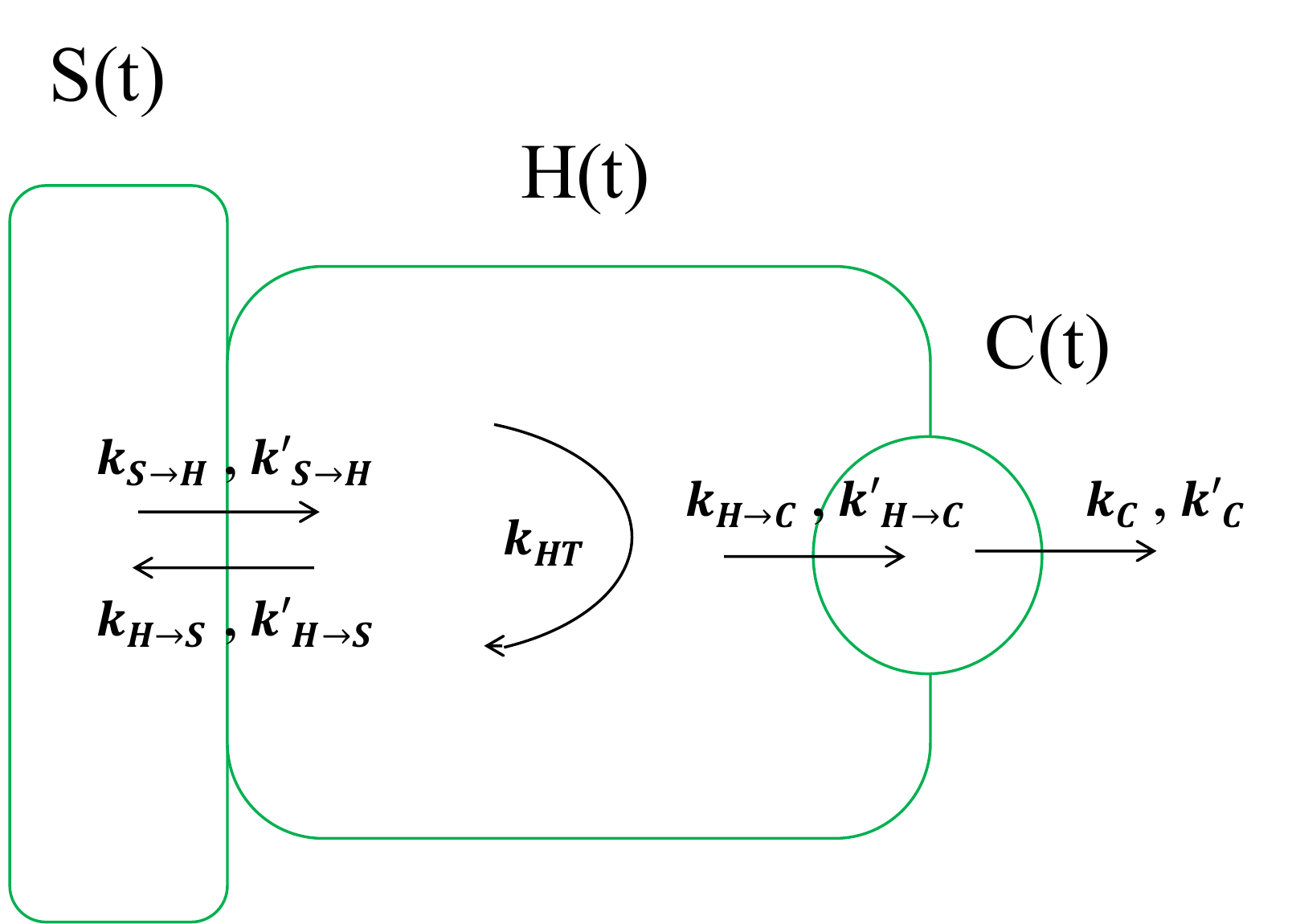}
\caption{\textbf{Schematic of Hepatic Transport Model Used}. We consider five transport rates for fluorescein: $k_{S\rightarrow H}$ for the transport from sinusoid to the hepatocyte, $k_{H\rightarrow S}$ from the hepatocyte back to the sinusoid, $k_{H\rightarrow C}$ from the hepatocyte to the canaliculus, and a loss rate $k_{C}$ from the canaliculus.  We also assume glucuronidated fluorescein is transported via the same mechanisms, albeit with rates that have different values, designated by the primed symbols above. Finally $k_{HT}$ represents the transformation of fluorescein into its glucuronidated form within the hepatocyte. The arrows show the direction of flow.}
\label{Schematic}
\end{center}
\vspace{-0.3in}
\end{figure}

A schematic of the model is provided in Fig.~(\ref{Schematic}). In our model, the species vector, previously written as $\textbf{x}$ in Eq.~(\ref{basicODE}), includes the unit of measurement for 
unmodified and modified (glucuronidated) fluorescein in each compartment.
The quantities are given by $S, H, C$ for fluorescein and $S', H', C'$, for glucuronidated fluorescein in each compartment. 
Finally, while we have six species, glucuronidated and unmodified fluorescein in three compartments, we only have three measurements, namely the fluorescence intensity in each compartment.

Based on the model schematic provided in Fig.~(\ref{Schematic}), after pre-specifying the input rate into the sinusoid thereby setting initial conditions, the dynamical model is given by

\begin{equation}
\left\{
 \begin{array}{ll}
\dfrac{dS}{dt}=-k_{S\rightarrow H}S+k_{H\rightarrow S}H \\ 
\\
\dfrac{dH}{dt}=k_{S\rightarrow H}S-(k_{H\rightarrow S}+k_{H\rightarrow C})H-k_{HT}H \\ 
\\
\dfrac{dC}{dt}=k_{H\rightarrow C}H-k_{C}C
 \end{array}
  \right.
\label{ODEunGF}
\end{equation}

and

\begin{equation}
\left\{
 \begin{array}{ll}
\dfrac{dS'}{dt}=-k'_{S\rightarrow H}S'+k'_{H\rightarrow S}H' \\ 
\\
\dfrac{dH'}{dt}=k'_{S\rightarrow H}S'-(k'_{H\rightarrow S}+k_{H\rightarrow C})H'+k_{HT}H \\ 
\\
\dfrac{dC'}{dt}=k'_{H\rightarrow C}H'-k'_{C}C'.
 \end{array}
  \right.
\label{ODEGF1}
\end{equation}

The measurement model is now
\begin{equation}
\begin{pmatrix} y_{S} \\ y_{H} \\ y_{C}\end{pmatrix} 
= 
\begin{pmatrix} \alpha & 0 & 0 & \alpha' & 0 & 0 \\ 0 & \beta & 0 & 0& \beta' & 0 \\  0 & 0& \gamma & 0 & 0& \gamma' \end{pmatrix}
\begin{pmatrix} S \\ H \\ C \\ S' \\ H' \\ C' \end{pmatrix} + \begin{pmatrix} \epsilon_1\\ \epsilon_2 \\ \epsilon_3 \end{pmatrix}.
\label{ODEGF2}
\end{equation}

The parameters $\alpha, \beta$, and $\gamma$ and their primes are our \textit{measurement parameters} 
for unmodified and glucuronidated fluorescein in each compartment. 
We note that, in this case, the measurement matrix $\textbf{H}$ is no longer square or diagonal; at any given time, we have fewer measurements than number of species in our model.

 Furthermore, just as we did with simulated data, we used the identifiability problem procedure (detailed in the Supplementary Information Appendix B) and, on this basis,
pre-specified the values for the measurement parameters of both fluorescein and glucuronidated fluorescein in the sinusoid, $\alpha = 0.5$ and $\alpha'=0.25$, and also the conversion rate between them in the hepatocyte, $k_{HT}=0.5$ $min^{-1}$. 

\subsection{Study on Sham Control and 5/6N Rat Model of Chronic Kidney Disease}

Here, we used IVM data from 5/6N rat models as these are often used as models for the study of chronic kidney disease~\cite{leblond2001downregulation}. 
To evaluate the functional outcomes of the 5/6N model on hepatic transport, we collected IVM data~\cite{ryan2014effects} in the liver of sham control operated rats Fig.~(\ref{ExpResults}A).
In this case, sham control operated rats were treated with the same anesthetic and surgical procedures without kidney removal (as opposed to 5/6N with kidneys removed).

The results for these studies  are shown in Tables~(\ref{RealDataShamResults}) and ~(\ref{RealData5-6NPresseResults}). 
Similar to previously published work, e.g.~\cite{ryan2014effects}, our results also show a meaningful change in hepatic transport in 5/6N as compared to the sham control. 
Concretely, our analysis reveals that the 5/6N, when compared to the  sham control operated rats, exhibited a  
decrease rate of hepatic uptake of fluorescein. Put differently, we anticipate differences in $k_{S\rightarrow H}$, $k_{H\rightarrow S}$, $k'_{S\rightarrow H}$, and $k'_{H\rightarrow S}$ between these two cases, as hepatic transport is impaired in the 5/6N rat.

\begin{table}[h]
	\centering
	\resizebox{\columnwidth}{!} {%
	\begin{tabular}{c c c c c c c c c c c c c}
		\hline\hline
		$k_{S\rightarrow H}$&$k_{H\rightarrow S}$&$k_{H\rightarrow C}$&$k_{C}$&$k_{S\rightarrow H}'$& $k_{H\rightarrow S}'$&$k_{H\rightarrow C}'$&$k_{C}'$&$\beta$&$\gamma$&$\beta'$&$\gamma'$ \\
		\hline\hline
		2.2440 & 2.1501 & 0.1726 & 0.3802 & 0.6744 & 2.5368 & 1.5433 & 0.9639 & 0.4731 & 0.1393 & 0.1340 & 0.3763 & Mean values\\
		0.1678 & 0.1082 & 0.1007 & 0.0963 & 0.1324 & 0.0751 & 0.1035 & 0.2076 & 0.0995 & 0.1179 & 0.0514 & 0.0985 & Standard deviation \\
		 \hline
	\end{tabular}
	}
	\caption{{\bf Experimental Results Using Sham Control Operated Rats.} Parameter estimates for the experimental data  Fig.~(\ref{ExpResults}A) (the rates are in units of $min^{-1}$).}
	\label{RealDataShamResults}
\end{table}

\begin{table}[h]
	\centering
	\resizebox{\columnwidth}{!} {%
	\begin{tabular}{c c c c c c c c c c c c c}
		\hline\hline
		$k_{S\rightarrow H}$&$k_{H\rightarrow S}$&$k_{H\rightarrow C}$&$k_{C}$&$k_{S\rightarrow H}'$& $k_{H\rightarrow S}'$&$k_{H\rightarrow C}'$&$k_{C}'$&$\beta$&$\gamma$&$\beta'$&$\gamma'$ \\
		\hline\hline
		2.0661 & 2.2067 & 0.1580 & 0.2753 & 0.5385 & 2.1169 & 1.4519 & 0.6943 & 0.6285 & 0.2305 & 0.5092 & 0.2564 & Estimated values\\
		0.1558 & 0.0592 & 0.0347 & 0.0997 & 0.1527 & 0.0372 & 0.2725 & 0.2617 & 0.0590 & 0.0218 & 0.0334 & 0.0125 & Standard deviation \\
		 \hline
	\end{tabular}
	}
	\caption{{\bf Experimental Results Using 5/6N Rat Models of Chronic Kidney Disease.} Parameter estimates for the experimental data  Fig.~(\ref{ExpResults}B) (the rates are in units of $min^{-1}$).}
	\label{RealData5-6NPresseResults}
\end{table}

To resolve model unidentifiability, we pre-specify $k_{HT}$ as well as the measurement parameters $\alpha$ and $\alpha'$ in our model. We chose $k_{HT}$ and those two parameters as their values are the easiest to determine via physiological experiments~\cite{grotte1985fluorescent, lee1985blood, bijsterbosch1981plasma} or via fluorescence lifetime imaging~\cite{hato2017two}. Our quantitative conclusions are insensitive to exact parameter estimates used initially for $k_{HT}$, $\alpha$ and $\alpha'$.

\subsection{Effect of Taurolithocholate}

In the previous subsection, we devised a control to assess the functional consequences of the 5/6N and recovered a change in transport rates from the sinusoid to the hepatocyte. 
Now, we look at different treatment controls using Taurolithocholate (TLC) treated rats~\cite{ryan2014effects}. TLC is a pharmaceutical agent that inhibits transport from the hepatocyte to the canaliculus and out from the canaliculus, so we expect these relevant rates to decrease.

TLC-induced cholestasis is a common experimental model for  drug-induced cholestasis~\cite{javitt1968effect, layden1977taurolithocholate, roma1994hepatic}. According to previous work, TLC impairs hepatic transport~ \cite{petzinger1994transport} and also significantly blocks hepatocyte uptake of sodium fluorescein~\cite{roma1994hepatic}. Thus, by using TLC treated rat models, we could evaluate our method to see how well it works in estimating transport rates from the hepatocyte to the canaliculus and transport rates from canaliculus out.

\begin{figure}[H]
\begin{center}
\hspace{-1.5cm}
     \includegraphics[scale=0.22]{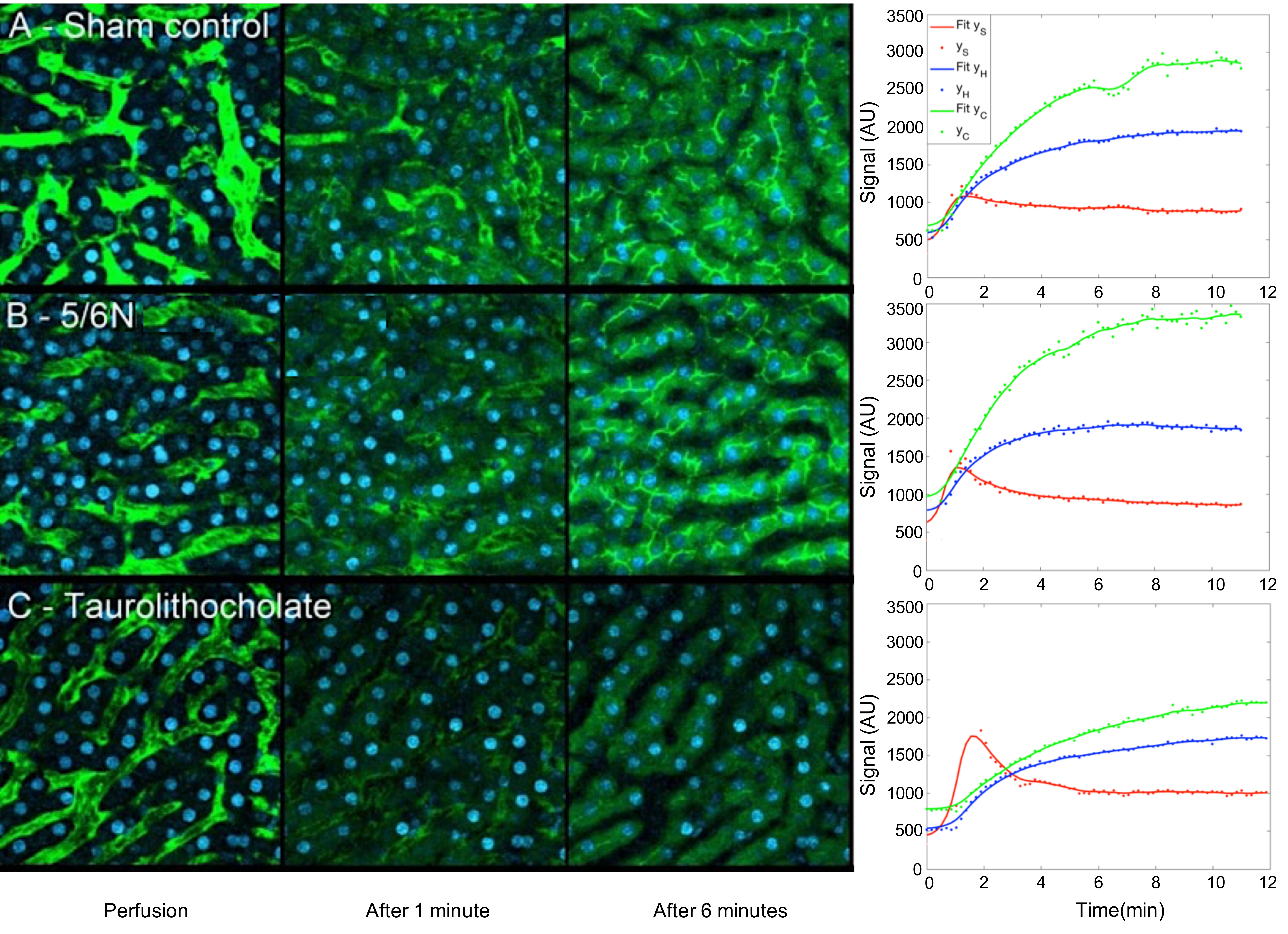}
\caption{{\bf Results of Studies of Fluorescein Transport in Rat Model Sham Control, 5/6N, and Taurolithocholate (TLC) Treated} Livers. In (A) we have sample images from a set of collected images from the liver of a rat as the sham control (without removing the kidneys) during intravenous injection of sodium fluorescein. On the right hand side we see our method's fit to the data with kinetic parameters responsible for the fit reported in Table~\ref{RealDataShamResults}. In (B) we show results for a 5/6N rat model in which the kidneys were removed. 
The main difference between (A) and (B) is the change in the rate of hepatic uptake of fluorescein as quantified by the rates from the sinusoid to the hepatocyte between the sham control and 5/6N rats for both glucuronidated and unglucuronidated forms of fluorescein.
More details on the rat model are provided in Ref.~\cite{ryan2014effects}. In (C) we show example images collected from a rat treated by the agent TLC obtained after intravenous injection. The TLC highly reduces the rate of fluorescein uptake into the canaliculus.} 
\label{ExpResults}
\end{center}
\vspace{-0.3in}
\end{figure}

The result of blocking hepatocyte uptake of sodium fluorescein using TLC treated rat on hepatocyte is shown in Fig.~(\ref{ExpResults}C). The estimated ODE parameter values for this data set appear in Table ~(\ref{RealDataBlockResults}) where we note 
the blocking effect TLC has on secretions to and from the canaliculus recovered by our model as measured by the small values for the rates $k_{H\rightarrow C}$ and $k_{C}$.

\begin{table}[h]
	\centering
	\resizebox{\columnwidth}{!} {%
	\begin{tabular}{c c c c c c c c c c c c c}
		\hline\hline
		$k_{S\rightarrow H}$&$k_{H\rightarrow S}$&$k_{H\rightarrow C}$&$k_{C}$&$k_{S\rightarrow H}'$& $k_{H\rightarrow S}'$&$k_{H\rightarrow C}'$&$k_{C}'$&$\beta$&$\gamma$&$\beta'$&$\gamma'$ \\
		\hline\hline
		1.6657 & 0.9180 & 0.0032 & 0.0026 & 3.3125 & 1.8065 & 1.0885 & 0.8773 & 0.2158 & 0.7079 & 0.1042 & 0.7474 & Estimated values\\
		0.0212 & 0.0332 & 0.0008 & 0.0005 & 0.0695 & 0.0800 & 0.0158 & 0.0286 & 0.0172 & 0.0502 & 0.0112 & 0.0407 & Standard deviation \\
		 \hline
	\end{tabular}
	}
	\caption{{\bf Experimental Results Using TLC.} The results of parameter estimation for the experimental data  Fig.~(\ref{ExpResults}C) (the rates are in units of $min^{-1}$).}
	\label{RealDataBlockResults}
\end{table}

The $k_{HT}$, $\alpha$, and $\alpha'$ were prespecified for the same reasons as for the Sham Control and 5/6N Rat cases above.


\section{Discussion and Conclusions}
\label{sec:discuss}

Drug development is a long and costly endeavor; the average drug costs nearly a billion dollars and takes roughly 15 years to bring to market~\cite{chong2007new, dimasi2003price}. Given these costs and timescales, it is critical to identify the efficacy and risks associated with a candidate drug early in the development process.  Clearly improving the prediction of drug failures could substantially reduce development costs~\cite{dimasi2003price, paul2010improve}. The need for improved tools for preclinical evaluation of drugs is the central focus of the FDA's \textit{Critical Path Initiative}~\cite{karsdal2009biochemical}. Although new drugs are scrutinized for effects on liver function, adverse effects on the liver comprise the most common biological reason for drug failure in the development of new pharmaceuticals~\cite{larrey2002epidemiology, larrey2000drug} and the most common cause for withdrawal of drugs from the market~\cite{giacomini2010membrane}.  The failure to predict these problems reflects fundamental shortcomings in the methods that are used in preclinical drug studies.

Our long-term goal is to combine mathematical modeling with IVM experimental data to determine the effects of drugs on hepatic transport.  The theoretical framework 
we develop here provides more accurate and reproducible measures of transport, including pathways that cannot be observed by other methods, supporting more powerful studies of {\it in vivo} liver function. 
By specifically addressing problems tied to fluorescence measurement,
our approach could increase the physiological relevance of {\it in vivo} studies in ways that could impact preclinical evaluations of the hepatic drug effects thereby extending the predictive power of {\it in vitro} drug development studies, minimizing the numbers of animals needed for {\it in vivo} studies and reducing the number of drug failures. As a first step towards developing new methods for the estimation of {\it in vivo} transport rate parameters, we have presented an implementation of the known method of parameter cascades for ODE parameter estimation, one that we tailored to IVM experimental data on hepatic transport.

In the context of Biophysics, parameter inference methods have a comparatively long history~\cite{sun2012parameter, chen2003bayesian, presse2013extracting, presse2011modeling, lee2012derivation, castillo2016flowcal, hebisch2013high, firman2017building, firman2018maximum}. 
The goal of parameter estimation is to find unknown parameters of the model that give the best fit to a set of experimental data~\cite{gabor2015robust}. 
While a number of methods tailored to learning parameters from ODEs exist, many of them require that the ODEs be numerically solved~\cite{bard1974nonlinear, biegler1986nonlinear} which entails expensive computation and requires knowing the initial values of the ODE variables. However, efficient computational methods exist that do not require actually solving the ODEs numerically~\cite{ramsay2006functional, brunel2008parameter, chen2008efficient}. 
A drawback for many of these methods is that they do not take into account errors approximation when making parameter inferences, which causes the well-known bias problem~\cite{conrad2015probability}. On the other hand, we deal with these problems through parameter cascades by defining two nested levels of optimization in our adaptation. In the inner optimization loop, 
we estimated nuisance parameters (coefficients of basis function). Then structural parameters are estimated in the outer optimization loop. 

Disadvantages of our method include the fact that weight assigned to the penalty term (the regularization parameter) can impact 
overall inference if unreasonable values are selected~\cite{cao2009generalized}. 
This is true for any Bayesian inference problem as well if unusual hyperparameters are selected~\cite{welch1995introduction} 
Furthermore, we only determine point estimates, rather than full posterior distributions, over the unknown parameter values~\cite{hines2015primer, hines2014determination, epstein2016bayesian, barenco2006ranked, brown2003statistical, vyshemirsky2008bayesian, sgouralis2018bayesian, sgouralis2017introduction}.

Regarding other approaches which focus on parameter estimation 
such as maximum likelihood and Bayesian mehods~\cite{tsekouras2016novel, chen2014molecular}, naive implementations demand that ODEs be solved first~\cite{epstein2016bayesian, chkrebtii2016bayesian, vyshemirsky2008bayesian}. Here with parameter cascades, this step is unnecessary even for highly non-linear dynamics (as exemplified by the FitzHugh-Nagamo results). 
On account of its ability to deal with non-linear dynamics as well as measurement parameters,
our method should be general enough to deal with  non-linearities introduced, say, by having kinetics dictated by  Michaelis-Menten ODE forms for all reactions. The latter would be especially relevant to capturing transporter saturation if such information is discernible from the data.

\section{Author Contributions}
\noindent  MT developed computational tools and analyzed data;
RD, and KWD contributed experimental data;
MT, KT, and SP conceived research;
SP oversaw all aspects of the projects.

\begin{suppinfo}

The following files are available free of charge.
\begin{itemize}
  \item Supplementary Information.pdf: A detailed description of the methods developed.
\end{itemize}

\end{suppinfo}

\begin{acknowledgement}
We acknowledge IU Collaborative Research Grants (IUCRG) for partially financial support. SP also acknowledges support from the NIH NIGMS  (R01GM130745-01).  We also thank the Indiana Center for Biological Microscopy for providing experimental data for the study.
\end{acknowledgement}

\begin{tocentry}
\includegraphics[width=0.7\textwidth]{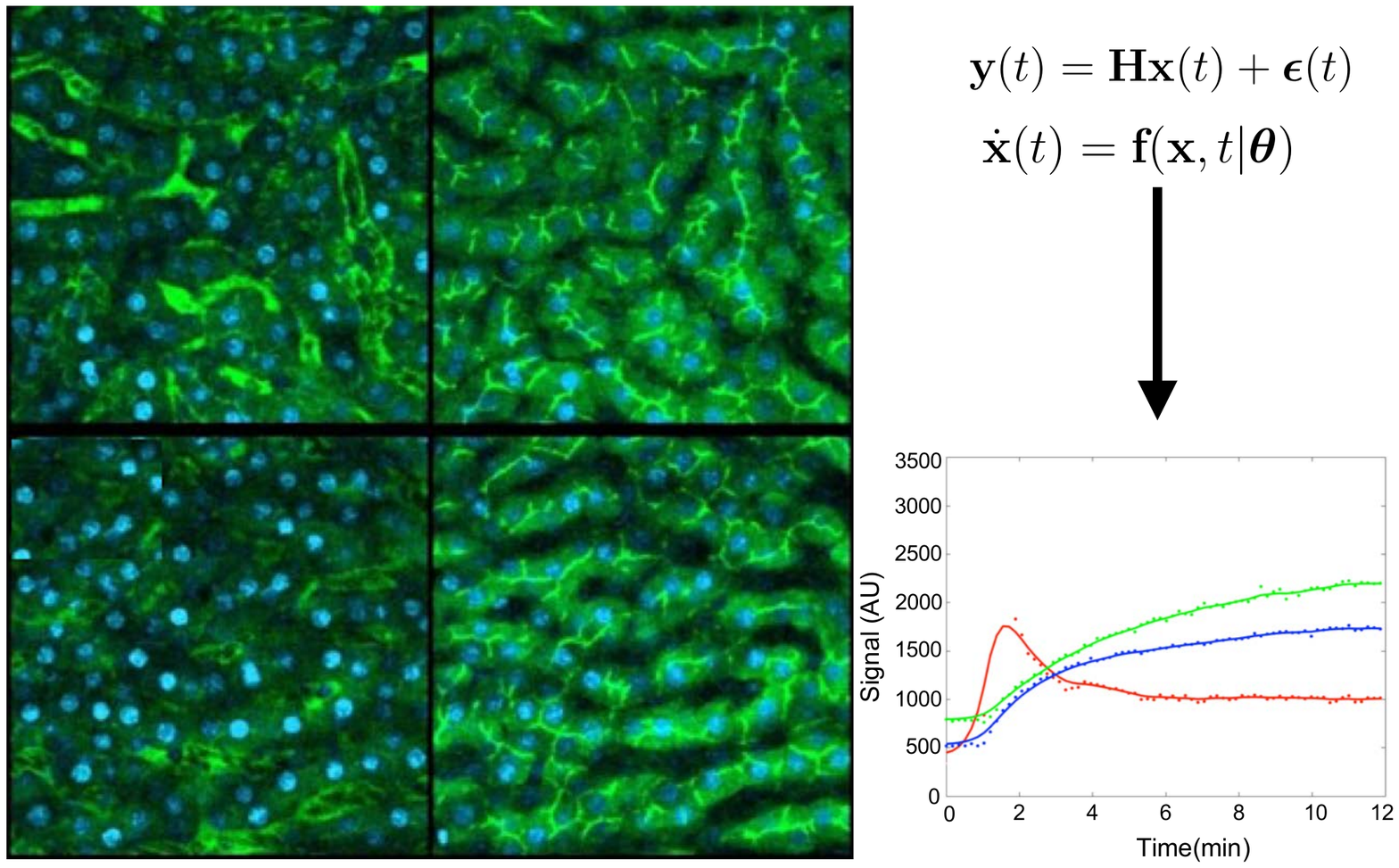}
\end{tocentry}

\bibliography{paper}

\end{document}